\documentclass{article}
\usepackage[a4paper,top=4.5cm,left=4cm,right=4cm,bottom=3.5cm]{geometry}
\usepackage{graphicx}

\usepackage{authblk}

\usepackage{url}
\usepackage[utf8]{inputenc}
\usepackage[dvipsnames]{xcolor}
\usepackage{proof}
\usepackage{tikz}
\usepackage{amssymb,amsmath,amsthm}
\usepackage{enumitem}
\usepackage[tableaux]{prooftrees}

\newcommand{\comment}[1]{}

\newcommand{\tuple}[2]{\langle #1, #2\rangle}
\newcommand{\truple}[3]{\langle #1, #2, #3\rangle}
\newcommand{\set}[1]{\{ #1 \}}

\newcommand{\biimp}{\leftrightarrow}
\newcommand{\imp}{\rightarrow}

\newcommand{\symdiff}{\,\triangle\,}
\newcommand{\cons}{\vdash}
\newcommand{\closure}{\divideontimes}
\newcommand{\absurd}{\divideontimes}
\newcommand{\exclusion}{\mathrel{\rangle\kern-0.18em\langle}}

\newcommand{\Assert}[1]{{\color{blue}{+}}\,{#1}}
\newcommand{\Deny}[1]{{\color{red}{-}}\,{#1}}

\newcommand{\Elim}[1]{\mathbf{E}_{#1}}
\newcommand{\AssertRule}[1]{{+}{#1}}
\newcommand{\DenyRule}[1]{{-}{#1}}

\newcommand{\highlight}[2]{\colorbox{#1}{$\displaystyle #2$}}

\definecolor{mygray1}{gray}{0.4}
\definecolor{mygray2}{gray}{0.35}

\newcommand{\hypnode}[2]{{
\scalebox{0.85}{\textcolor{mygray2}{$(#1)$ }}}
#2
}
\newcommand{\node}[3]{
\scalebox{0.85}{{\textcolor{mygray2}{$(#1)$ }}}
#2
\scalebox{0.85}{{\textcolor{mygray2}{\, $#3$}}}
}
\newcommand{\close}[1]{\raisebox{1ex}{\textcolor{mygray1}{\scalebox{0.75}{#1}}}}

\theoremstyle{definition}
\newtheorem{theorem}{Theorem}

\newtheorem{example}[theorem]{Example}

\colorlet{LightRed}{Red!90}

\usepackage{setspace}

\usepackage{caption}
\usepackage[backend=biber, style=numeric]{biblatex}
\usepackage[hidelinks]{hyperref} 

\usepackage{titlesec}
\titleformat{\section}
  {\fontsize{11}{11}\selectfont\bfseries}
  {\thesection.}
  {0.3em}
  {}

\titleformat{\subsection}
  {\fontsize{11}{11}\selectfont\bfseries}
  {\thesubsection.}
  {0.3em}
  {}

\title{\textbf{A method for the automated generation of proof exercises with comparable levels of proving complexity}}

\author[1]{João Mendes}
\author[2]{João Marcos}
\author[3]{Patrick Terrematte}

\affil[1]{{\normalsize Graduate Program in Informatics, University of Brasília}}
\affil[2]{{\normalsize Philosophy Department, Federal University of Santa Catarina}}
\affil[3]{{\normalsize Metrópole Digital Institute, Federal University of Rio Grande do Norte}}

\date{}

\begin{document}

\maketitle

\begin{abstract}
The automated generation of exercises may substantially reduce the time educators devote to manual exercise design.
A major obstacle to the integration of such automation into teaching practice, however, lies in the ability to control the difficulty of mechanically generated exercises. 
This paper presents a method for the automated generation of proof exercises with comparable levels of proving complexity. The method takes as input a proof exercise together with a set of rules that yield a proof of the exercise, and produces as output a set of proof exercises whose proving complexity is comparable to that of the input exercise.
The approach focuses on mathematical proof exercises formulated in first-order languages, covering topics typically addressed in undergraduate Discrete Mathematics courses. 
We assess the proving complexity of these exercises by considering the effort required to solve them through informal proofs, 
and argue that this effort can be formally captured through cut-based tableau proofs that are free of logical symbols. 
The rules governing such proofs are obtained through a mechanizable extraction procedure introduced in this paper. 
By exploiting the analytic nature of these rules and the structure inherent in proofs constructed via tableau rules, we derive a computational procedure implementing the proposed method.

\end{abstract}

\section{Introduction}\label{sec:intro}

Educators routinely face substantial demands beyond classroom teaching, including lesson planning, providing extracurricular student support, assigning exercises, and designing assessments.  Among these tasks, the manual construction of exercises is particularly time-consuming and represents a significant component of instructional workload.  This challenge motivates the development of Automatic Question Generation (AQG) tools, which aim to reduce the effort required to produce educational materials.  AQG systems generate questions from knowledge sources that may be structured (e.g., databases and ontologies) or unstructured (e.g., continuous text), while allowing educators to specify desired question properties, such as the number of distractors in multiple-choice questions \cite{Kurdi2020}.  Despite these advances, ensuring that automatically generated exercises meet pedagogical requirements remains a central challenge.

The lack of AQG tools supporting fine-grained control over proving complexity remains a major obstacle to their meaningful adoption in pedagogical settings \cite{AlFaraby2024, Alsubait2015, Kurdi2020}.  Addressing this limitation, we propose a method for the automated generation of proof exercises with comparable levels of proving complexity.  By \textit{proof exercise}, we refer to exercises whose statements contain provable conjectures.  The proving complexity of such exercises is assessed through a theory-specific tableau-based framework introduced in this paper.

A key property of theory-specific tableaux is that the formulas occurring in their proofs contain no logical symbols.  The rules used to construct these tableaux are extracted through a mechanizable procedure from certain closed first-order axioms.  These axioms are designed to characterize fragments of first-order mathematical theories typically taught in undergraduate Discrete Math courses, such as Set Theory and Number Theory, thereby enabling the systematic derivation of provable conjectures suitable for classroom proof exercises.

Theory-specific tableaux can be understood as an instance of the so-called ``KE methodology'', which we here refer to as a \textit{cut-based approach}, originally developed by D'Agostino and Mondadori for classical logic \cite{DAgostino1990, DAgostino1994}.
A central feature of this methodology is that cut-based tableaux can polynomially simulate truth-table procedures, whereas Smullyan-style tableaux do not enjoy this property \cite{DAgostino1992}. 
Such computational advantage stems from restricting branching to applications of an analytic cut rule, thereby providing tighter control over proof expansion. 
In analytic cuts, the cut formula is typically required to be a subformula of a formula already present in the tableau. 
In fact, building on this line of investigation, generalizations of the subformula property, and of analyticity more broadly, have been explored in \cite{Caleiro2015}, whose results motivate and inform the developments introduced in the present work.

This paper is organized as follows. Section \ref{sec:related} reviews existing approaches to difficulty control in AQG and identifies the gaps addressed by our method.  Section \ref{sec:ThBg} introduces the syntactic, deductive, and semantic notions employed throughout the paper.  The procedure for extracting the rules used to construct theory-specific tableaux is presented in Section~\ref{sec:Tsproofs}.  
Next, section \ref{sec:complexity_proof_exercises} defines the notion of proving complexity and presents the pedagogical motivations for employing theory-specific tableaux to determine it.
The procedure for generating exercises with comparable levels of proving complexity is described in Section~\ref{sec:main_procedure_description}.  Finally, Section \ref{sec:conclusion} offers concluding remarks and outlines directions for future work.

\section{Related work}\label{sec:related}

Several literature reviews on AQG have already pointed out the lack of AQG tools endowed with difficulty control. In \cite{Kurdi2020}, the authors conducted a review on 93 papers about AQG for educational purposes published from 2015 to 2019. From these 93 papers, only 14 presented methods that had a mechanism for controlling the difficulty of their outputs. As posed in \cite{Alsubait2015}, this challenge is not a novelty since the scarcity of approaches for the generation of questions with controlled levels of difficulty was also identified in a similar analysis of 81 papers published between 1969 and 2015. And a more recent study, \cite{AlFaraby2024}, comprising AQG tools that use deep learning for educational purposes, highlights the customization of these tools by difficulty level as ``a further research opportunity''.

From a pedagogical perspective, controlling difficulty is a key feature for AQG tools. First, it guarantees that the exercises are produced in accordance with the expected proficiency from their test takers. 
Second, AQG tools equipped with difficulty-control mechanisms can support tutors in constructing personalized formative assessments ---in other words, assessments delivered during a course with the goal of providing ongoing feedback that matches each student’s level of mastery. Such tools can also assist in the development of computerized adaptive tests, that is, dynamic personalized assessments that select subsequent questions based on the test-taker’s performance on previous ones.
And third, for individualized assessments, difficulty control avoids benefiting some students by assigning to them easier questions than to their colleagues.

In spite of the lack of AQG techniques with difficulty control, some domain-independent approaches have already been proposed in the literature. Question Difficulty Estimation from Text (QDET) \cite{Benedetto2023, AlKhuzaey2024} is a recent growing trend in Natural Language Processing in order to automate the process of estimating the difficulty of questions extracted from texts. In QDET, the usual approach is to provide a machine learning model that predicts the difficulty of a question using a previous classification made by humans.

One of the drawbacks of QDET, and machine learning models in general, is that classifying a question according to human judgements may lead to inconsistencies, since even experts can disagree, for example, on whether a question is difficult to be answered by a student or not. Another drawback is that the classification of difficulty provided by these models may not come accompanied by an explanation. As it is going to be detailed in Sections~\ref{sec:complexity_proof_exercises} and~\ref{sec:main_procedure_description}, these drawbacks do not affect our method since it does not depend on any previous human judgment and the classification of difficulty of the proof exercises is explained by their minimal proofs.

Regarding tools that are more closely related to our scope and that provide explainable mechanisms to control the difficulty of questions, a metric employed to classify the difficulty of a logic-related proof exercise is its syntactic structure, as in \cite{Patrick2013,Terrematte2011}. 
In these studies, the authors present a method for generating formulas of propositional logic in which the proving complexity of the resulting exercises is regulated through parameters such as the number of propositional atoms, the number of connectives, and the depth of the formulas. These parameters are measured with respect to the abstract representation of formulas, rather than their concrete representation. While the concrete representation typically expresses formulas as sequences of symbols, the abstract representation makes explicit the internal structure of the mathematical object. For instance, the abstract representation of a propositional formula~$\varphi$ can be modeled as a tree whose root corresponds to the main connective of~$\varphi$.
Although the selection of such parameters may be justified by their impact on the size of the associated truth tables, this line of reasoning does not readily extend to first-order languages.
More importantly, two provable propositional formulas $\varphi_1$ and $\varphi_2$ may have the same syntactic structure, but the structure of the proofs for them may be very different. So, students might have more difficulty in proving $\varphi_1$ than in proving $\varphi_2$, or vice versa. To control the difficulty of the generated exercises, the method proposed in \cite{Patrick2013,Terrematte2011} only takes into account factors that precede and are not necessarily influenced by the solution(s) of the produced exercises.

Similarly, in \cite{Singh2021}, the authors used the syntactic structure of proof exercises about Algebra involving equational reasoning to produce other proof exercises under this same scope with comparable levels of difficulty. This system also allows the user to add constraints during the process of generation as a mechanism to control difficulty. For example, if the input exercise is an equality that has an irreducible fraction in its right-hand side, the user can constrain the method to generate only equality exercises whose right-hand sides are irreducible fractions. In some cases, the authors argue that the exercises generated by their method may be proved using ``similar proof strategies''. However, no further detail on this is provided. Again, as the syntactic structure of an exercise described with a formal language does not necessarily determine the size of a solution for it, the difficulty level assigned to an exercise may fail to capture the actual effort involved in solving it. 

In a sense, our approach is more closely related to the one presented in \cite{Wang2016}, where the proposed method generates Mathematical Word Problems ---that is, exercises formulated as short narratives containing unknown values to be determined using only the information provided in the text---, whose difficulty level can be controlled by varying the number of arithmetic operations required to solve the corresponding equation.
The results of an experiment conducted with a group of 30 students revealed that: increasing the amount of operations in an exercise increases its error rate, indicating that it becomes more difficult; and exercises with the same amount of operations had indistinguishable error rates. 
Although these results concern a different type of exercise, they indicate that using the structure of the solution as a basis for characterizing exercise difficulty may provide a more promising direction within formal disciplines such as Logic.

\section{Preliminaries}\label{sec:ThBg}

Our method is designed to handle proof exercises formulated in first-order languages. As usual, a \textit{first-order language} is specified over a first-order signature, consisting of a family of function symbols and a family of predicate symbols, and a denumerable set of variables. The propositional connectives are $\bot, \top, \lnot, \land, \lor, \imp$ and $\biimp$, and the first-order quantifiers are $\forall$ and $\exists$. An \textit{atomic formula}, also called \textit{theory-specific formula}, does not contain quantifiers nor connectives. A \textit{literal}, as usual, is either an atomic formula or a nullary connective, or the negations thereof, and a \textit{clause} is a disjunction of literals. A literal is said to be \textit{positive} if it is an atomic formula and it is said to be \textit{negative} otherwise. The \textit{opposite of a positive literal}~$l$ is $\lnot l$ and the \textit{opposite of a negative literal} $\lnot l$ is $l$. We denote the opposite of a literal~$l$ by~$l^\text{op}$.
The \textit{abstract representation} of a theory-specific formula $sf$ is given by a tree whose root is the predicate symbol of $sf$. For any other node labeled by a function symbol~$f$, its $i$-th children corresponds to the $i$-th argument of~$f$ in $sf$. The \textit{depth} of a theory-specific $sf$ formula is equal to height of its abstract representation.

A \textit{signed formula} is a formula preceded by either a ``$\color{blue}{+}$'' sign or a ``$\color{red}{-}$'' sign. For a signed formula $sf$, we denote its underlying formula by $fmla(sf)$ and its sign by $sign(sf)$. The \textit{conjugate} of a signed formula $sf_1$ is a signed formula $sf_2$ such that $fmla(sf_1)=fmla(sf_2)$ and $sign(sf_1)\not=sign(sf_2)$.

Analytic tableaux \cite{Smullyan1968} are deductive systems in which proofs are represented as trees whose nodes are labeled by formulas. These trees are expanded using \textit{tableau expansion rules}. A tableau expansion rule is defined as a schema, consisting of a premise set and a conclusion set, representing all rule instances generated from this schema. A rule~$r$ is applicable to a tableau~$\texttt{T}$ if, for some branch~$\theta$ of~$\texttt{T}$, a set of formulas occurring in~$\theta$ matches instances of the premises of~$r$. The rule is applied by expanding the branch~$\theta$ with the corresponding instances of the alternative conclusions of~$r$. The application of a rule may increase the amount of branches in the tableau. In this study, we primarily consider \textit{linear expansion rules}, that is, rules whose application do not introduce additional branches. Due the procedure presented in Section \ref{sec:Tsproofs}, these linear rules are single conclusion.

Let $SF$ be a finite set of signed formulas and $ER$ be a set of tableau expansion rules. A branch whose nodes are labeled exactly by the signed formulas in $SF$ constitutes an $ER$\textit{-tableau} for $SF$ and, if $\texttt{T}$ is an $ER$-tableau for $SF$ and $\texttt{T}^*$ is obtained by applying any rule in $ER$ to $\texttt{T}$, then $\texttt{T}^*$ is also an $ER$-tableau for $SF$. A branch in a $ER$-tableau is \textit{closed} if it contains either $\Assert\!{\bot}$ or $\Deny\!{\top}$ or a signed formula and its conjugate, and it is \textit{open} otherwise. Closed branches are marked with the symbol $\closure$ below them. These notions extend naturally to entire $ER$-tableaux: an $ER$-tableau is \textit{closed} if all its branches are closed and it is \textit{open} otherwise.

An $ER$\textit{-proof} of a set of formulas $\Delta := \set{\delta_1, \dots, \delta_n}$ from a set of formulas $\Gamma := \set{\gamma_1, \dots, \gamma_m}$ is a closed $ER$-tableau for $P := \set{\Assert \gamma_1, \dots, \Assert \gamma_m, \Deny \delta_1,$ $\dots, \Deny \delta_n}$. We say that $\Delta$ is $ER$\textit{-provable} from $\Gamma$ if there is an $ER$-proof of $\Delta$ from $\Gamma$. We denote this by $\Gamma \cons_{ER} \Delta$ and refer to it as the \textit{consequence relation induced by} $ER$. Moreover, the set $P$ is said to be $ER$\textit{-refutable}.

A node $n$ in a tableau $\texttt{T}$ is a \textit{descendant} of a node $m$ whenever $n$ is added to $\texttt{T}$ as the result of applying an expansion rule whose premises include either the signed formula in~$m$ or in a node $m^\prime$ that is itself a descendant of~$m$. In the former case, $n$ is called a \textit{direct descendant of}~$m$. In a branch $\theta$ of tableau, nodes whose signed formulas are conjugate are referred to as the \textit{closure nodes} of $\theta$. A tableau is said to be \textit{clean} if every node in it is a descendant of some closure node. As will be shown in Section \ref{sec:main_procedure_description}, the brute-force procedure for constructing proofs from a set of signed formulas may naturally generate tableaux that are not clean. Such tableaux will not be relevant for our purposes.

The notions of satisfiability and validity we adopt are the standard ones for classical first-order logic. Additionally, for signed formulas, a model is said to satisfy $\Assert{\varphi}$ if it satisfies~$\varphi$ and is said to satisfy $\Deny{\varphi}$ if it does not satisfy~$\varphi$. Two (signed) formulas are said to be \textit{equisatisfiable} if they are satisfied by the same models; the definition extends naturally for the case of sets of (signed) formulas.
A signed formula $sf$ such that $fmla(sf)=\varphi$ is transformed into an equisatisfiable unsigned formula by the function $eq\_fmla$ defined as $eq\_fmla(sf)=\varphi$ if $sign(sf)=\Assert{}$ and $eq\_fmla(sf)=\lnot\varphi$ otherwise.

Normal forms are used to transform formulas into equisatisfiable formulas having a specific format. Here, we make use of a normal form that we call \textit{rule implicational normal form} (RINF). Formulas in RINF have the form $(l_1\land\dots\land l_{n-1})\imp l_n$, where $l_1,\dots, l_n$ are literals. By using valid equivalences in classical logic, such as De Morgan's laws and the Replacement Theorem, it is possible to verify that every formula in propositional classical logic can be put into RINF. 
Also, a set of signed formulas $\Gamma$ is in \textit{signed theory-specific normal form} (STSNF) if all formulas in $\Gamma$ are signed theory-specific formulas. Do note that some sets of signed formulas cannot be converted into an equisatisfiable set in STSNF. 
For example, the set $S := \{\Assert{}\varphi_1\lor\varphi_2\}$ cannot be put into STSNF as any set $S^\prime$, where $sf\in S^\prime$ only if $fmla(sf)\in\{\varphi_1,\varphi_2\}$, is equisatistiable with $S$.

A tableau is said to be \textit{satisfiable} if at least the set of all signed formulas of one of its branches is satisfiable. The soundness of a tableau system for classical logic requires that, should a rule be applied to a satisfiable tableau, the resulting tableau remains satisfiable. This implies, for a linear expansion rule~$r$, that the (single) conclusion of~$r$ is satisfiable if the premises of~$r$ are. It follows that a linear expansion rule~$r$ having a set~$SF$ of signed formulas as its premises and a signed formula~$sf$ as its conclusion is sound if $\varphi:=\bigwedge\limits_{s\in SF} eq\_fmla(s) \imp eq\_fmla(sf)$ is satisfiable. Do note that $\varphi$ is in RINF. In this case the rule $r$ is said to be the \textit{correspondent rule} of $\varphi$ and~$\varphi$ is the \textit{correspondent formula} of~$r$. Accordingly, we may say that a rule is satisfiable if its correspondent formula is.

\section{Theory-specific proofs}\label{sec:Tsproofs}

We intend to calculate the proving complexity of an exercise with the help of cut-based tableau proofs whose nodes are labeled by theory-specific signed formulas, and to attain that goal some definitions are in order. In particular, we shall refer to such tableau proofs as \textit{theory-specific proofs} and to the rules governing them as \textit{theory-specific rules}.

Theory-specific rules are extracted from definitional theories. A \textit{definitional theory}, in turn, is a structure $\truple{L}{Ax}{<}$, where $L$ is a first-order language, $Ax \subseteq L$ is a finite set of closed formulas not containing $\bot$ nor~$\top$, which we will refer to as \textit{definitional axioms}, and $<$ is a well-founded relation on the predicate symbols of~$L$.

\begin{example}
Here, we introduce a definitional theory that we will use as a running example throughout the paper: the definitional theory of sets, or \textit{theory of sets} for short. It is designed to allow us to derive a fragment of the theorems from Set Theory. The language of the theory of sets is defined by the signature $\Sigma := \tuple{FS}{PS}$, in which:

\begin{itemize}
\item The only constant symbol (nullary function symbol) of $\Sigma$ is ``$\emptyset$''.
\item The only unary function symbols of $\Sigma$ are ``$\complement$'', ``$f\!st$'' and ``$snd$''.
\item The only binary function symbols of $\Sigma$ are ``$\cup$'', ``$\cap$'', ``$\setminus$'', ``$\times$'' and ``$\triangle$''.
\item $FS_i$ is empty for any $i > 2$.
\item The only binary predicate symbols of $\Sigma$ are ``$\in$'', ``$\subseteq$'', and ``$\exclusion$''.
\item $PS_i$ is empty for any $i \not= 2$.
\end{itemize}

For binary symbols, we adopt the infix notation and omit parentheses when there is no chance of ambiguity. The set of definitional axioms $Ax$ is given by the formulas presented below:

\begin{itemize}
\item $[ax\emptyset]$ Definitional axiom for $\emptyset$: $(\forall x) (\neg x \in \emptyset)$.
\item $[ax\complement]$ Definitional axiom for $\complement$: $(\forall x, y) (x \in \complement(y) \biimp \neg x \in y)$.
\item $[ax\cup]$ Definitional axiom for $\cup$: $(\forall x, y, z) (x \in y \cup z \biimp (x \in y \lor x \in z))$.
\item $[ax\cap]$ Definitional axiom for $\cap$: $(\forall x, y, z) (x \in y \cap z \biimp (x \in y \land x \in z))$.
\item $[ax\setminus]$ Definitional axiom for $\setminus$: $(\forall x, y, z) (x \in y \setminus z \biimp (x \in y \land \neg x \in z))$.
\item $[ax\times]$ Definitional axiom for $\times$: $(\forall x, y, z) (x \in y \times z \biimp (f\!st(x) \in y \land snd(x) \in z))$.
\item $[ax\triangle]$ Definitional axiom for $\triangle$: $(\forall x, y, z) (x \in y \symdiff z \biimp ((\neg x \in y \imp x \in z) \land (x \in y \imp \neg x \in z)))$.
\item $[ax\subseteq]$ Definitional axiom for $\subseteq$: $(\forall x, y) (x \subseteq y \biimp (\forall z) (z \in x \imp z \in y))$.
\item $[ax\exclusion]$ Definitional axiom for $\exclusion$: $(\forall x, y) (x \exclusion y \biimp (\forall z) \neg(z \in x \land z \in y))$.
\end{itemize}

The only definitional axiom in which the unary symbols $f\!st$ and $snd$ are present is $[ax\times]$. Intuitively, we want $x \in y \times z$ to be the case when $x$ is a pair $(a, b)$ such that $a \in y$ and $b \in z$. To avoid axioms concerning pairs, we use $f\!st$ to allow for the first element of a pair to be recovered and we use $snd$ to recover the the second element of a pair.
Finally, the well-founded relation $<$ of the theory of sets is defined by setting 
$x<y$
if and only if 
$x$ is the predicate symbol of membership (that is, $\in$) and $y$ is one of the other predicate symbols (that is, either~$\subseteq$ or~$\exclusion$).
\end{example}
\vspace{1em}

One could wonder why the definitional axiom for $\times$ is not the formula $(\forall x_1, x_2, y, z)$ $((x_1, x_2)\in y\times z\biimp (x_1\in y\,\land\, x_2\in z))$.  As explained in Section~\ref{sec:complexity_proof_exercises}, the notion of comparability between proofs adopted here is based on the isomorphism between proofs that do not contain logical symbols. Such proofs are constructed from rules extracted from the correspondent definitional axioms. By formulating definitional axioms whose atomic subformulas are themselves isomorphic, we increase the likelihood of obtaining isomorphic proofs.

Given a definitional theory $\truple{L}{Ax}{<}$, we want to be able to prove conjectures about the language~$L$ using the definitional axioms as base assumptions. In that sense, definitional axioms may be seem as 0-premise rules. However, for pedagogical reasons discussed in Section \ref{sec:complexity_proof_exercises}, the proofs had better not contain logical symbols. So, we need to extract rules that act just like these definitional axioms and that contain logical symbols neither in their premises nor in their conclusions. Before showing how such rules are extracted, we detail their structure.

In Smullyan-style tableaux \cite{Smullyan1968}, analyticity is typically understood in terms of immediate subformulas: a formula is analyzed compositionally through the decomposition of its direct constituents. Since theory-specific formulas are atomic, this standard notion of analyticity is insufficient for theory-specific proofs and must therefore be extended. In \cite{Caleiro2015}, a generalized notion of analyticity is proposed, in which the syntactic complexity of a formula~$\varphi$ is determined by formulas that are related to~$\varphi$ and may be said to be less syntactically complex than~$\varphi$, but that are not necessarily immediate subformulas of~$\varphi$. For theory-specific proofs, indeed, we follow a similar line of approach: the single conclusions of the rules are constructed with terms already occurring in the premise set, and reduction of syntactical complexity is ensured through the interaction between the well-founded relation on predicate symbols of a definitional theory and the syntactic depth of theory-specific formulas. In what follows, we define the linear expansion rules, which we shall call theory-specific rules, and describe the procedure by which they are extracted from the definitional axioms. From these linear expansion rules, a family of cut rules for theory-specific proofs is introduced, as instances of a general cut schema.

A \textit{theory-specific rule} of a definitional theory $Th := \truple{L}{Ax}{<}$ is a pair consisting of a set of theory-specific signed formulas (the premises) and a theory-specific signed formula (the conclusion) respecting the following three theory-specific analytic restrictions:
\begin{enumerate}
\item The variables in the conclusion must be present in the premise set.
\item The predicate symbol in the conclusion is smaller, according to the relation~$<$, than the predicate symbol in at least one of the premises.
\item If all predicate symbols in the premise set and in the conclusion are the same, then the depth of the formula in the conclusion is smaller than the depth of at least one of the premises.
\end{enumerate}

Due to the first restriction, each application of a theory-specific rule to a tableau~$\texttt{T}$ admits only finitely many instances, since the conclusion that is being added to~$\texttt{T}$ is constrained to have only symbols that are already present in~$\texttt{T}$. The second and third restrictions ensure that we are reducing in some way the formulas in the proof. Figure~\ref{fig:sets_rules} illustrates the application of three rule schemas from the theory of sets. In each case, the signed formulas above the bar represent the premises of the rules and the signed formula below the bar represents the conclusion. Among these, only schema (c) qualifies as theory-specific: schema (a) violates the first and third restrictions, while schema (b) violates the second restriction.

\begin{figure}[h]
\centering

\begin{tabular}{ccc}
\raisebox{0.5em}{
\begin{tabular}{c}
(a)
\\
$
\infer[]{\Assert{x\in y\cup z}}{\Assert{x\in y}}
$
\end{tabular}
}
\hspace{.5cm}
&
\begin{tabular}{c}
(b)
\\
$
\infer[]{\Deny{x\subseteq y}}{\deduce{\Deny{z\in y}}{\Assert{z\in x}}}
$
\end{tabular}
\hspace{.5cm}
&
\begin{tabular}{c}
(c)
\\
$
\infer[]{\Deny{z\in x}}{\deduce{\Assert{x\subseteq y}}{\Deny{z\in y}}}
$ 
\end{tabular}
\end{tabular}
\caption{Rule schemas for the definitional theory of sets.}
\label{fig:sets_rules}
\end{figure}

In classical logic, to show that the consequence relation induced by a set of rules $ER\cup \{r_1\}$ coincides with the consequence relation induced by a set of rules $ER\cup \{r_2\}$, it suffices to prove that~$r_1$ and~$r_2$ are equisatisfiable. For each definitional axiom, our strategy is to translate it into an equisatisfiable set~$\Gamma$ of formulas in the Rule Implicational Normal Form. This translation proceeds by first converting the axiom into Prenex Normal Form (PNF), then applying Skolemization, then transforming the result into Conjunctive Normal Form (CNF), and finally rewriting the resulting set of formulas in RINF.

Let $ax$ be a definitional axiom. We begin by converting $ax$ into PNF, skolemizing the resulting formula if necessary be, and then transforming its matrix ---that is, the quantifier-free part of the formula obtained after removing all the initial quantifiers--- into CNF. As a result, we obtain a set of clauses $\Gamma$ that is equisatisfiable with $ax$. Each clause $\varphi:=l_1\lor\dots\lor l_n$, containing $n$ literals with $n>1$, can be transformed into $n$ formulas in RINF, namely $\varphi_1:=(l_2^\text{op}\land\dots\land l_{n}^\text{op}) \imp l_1,$ $\dots,$ $\varphi_n:= (l_1^\text{op}\land\dots\land l_{n-1}^\text{op}) \imp l_n$, such that $\varphi$, $\varphi_1$, $\dots$, $\varphi_{n-1}$ and $\varphi_n$ are all equisatisfiable. The set of rules extracted from $ax$ consists in the correspondent rules to these formulas $\varphi_i$ in RINF, for $1\leq i\leq n$. These rules contain no logical symbols in either their premise sets or their conclusions and satisfy the theory-specific analytic restrictions.

A clause containing a single negative literal $\lnot l$ can be transformed into two equisatisfiable formulas in RINF, namely: $\varphi_1:=l\imp\bot$ and $\varphi_2:=\top\imp\lnot l$. The correspondent rule to $\varphi_1$ is $r_1:=\tuple{\set{\Assert{l}}}{\Assert{\bot}}$, which states that a branch may be closed whenever it contains the signed formula $\Assert{l}$. Accordingly, the rule $r_1$ is represented as $\tuple{\set{\Assert{l}}}{\closure}$ and added to the set of extracted rules. The correspondent rule to $\varphi_2$ is $r_2:=\tuple{\set{\Assert{\top}}}{\Deny{l}}$. This rule allows $\Deny{l}$ to be introduced in any branch at any stage, thereby yielding infinitely many instances. For this reason, $r_2$ is not included among the extracted rules. If the single literal on this clause is positive, the extracted rule is $\tuple{\set{\Deny{l}}}{\closure}$.

We could have adopted a more general normal form than RINF during the procedure of extraction, namely: $(l_1\land\dots\land l_{i})\imp (l_{i+1}\land\dots\land l_n)$, where $l_1,\dots, l_i,\dots, l_n$ are literals. In this case, we would able to extract linear rules with more than one conclusion. However, the amount of formulas in this general normal form that could be transformed a clause containing $n$ literals is in the order of $n!$. This is why we constrain theory-specific rules to be linear.

For the remainder of this paper, whenever we mention ``the set of theory-specific rules for a definitional theory $Th$'' we will be referring to the set of theory-specific rules extracted from the definitional axioms of $Th$ following the above recipe.

\begin{example}
Consider the definitional axiom $[ax\subseteq]$ from the theory of sets. The result of converting it into PNF and skolemizing it is $(\forall x, y, z)((f(x,y) \in x \imp f(x,y) \in y) \imp x \subseteq y) \land (x \subseteq y \imp (z \in x \imp z \in y))$, in which~$f$ is a skolem function symbol. Then, after performing the conversion of the matrix of this formula into CNF, we are left with the following set of clauses $\{f(x,y) \in x \lor x \subseteq y$, $\lnot f(x,y) \in y \lor x \subseteq y$, $\lnot x \subseteq y \lor (\lnot z \in x \lor z \in y)\}$. The set of theory-specific rules extracted from $[ax\subseteq]$, as well as from the other definitional axioms of the theory of sets, is presented in Figure \ref{fig:ts_sets_rules}. We refer to this set as $TS_{sets}$. Do note that the function symbol $g$ is also a skolem function symbol.
\end{example}

\begin{figure}[ht]
\centering

\begin{center}
\scalebox{0.85}{
\begin{tabular}{ccc}
$
\infer[\AssertRule{\emptyset}]{\closure}{\Assert{x\in\emptyset}}
$
&
$
\infer[\AssertRule{\complement}\Elim{}]{\Deny{x\in y}}{\Assert{x\in \complement(y)}}
$
&
$
\infer[\DenyRule{\complement}\Elim{}]{\Assert{x\in \complement(y)}}{\Deny{x\in y}}
$
\end{tabular}
}
\end{center}

\begin{center}
\scalebox{0.85}{
\begin{tabular}{cccc}
$
\infer[\AssertRule{\cup}\Elim{1}]{\Assert{x\in z}}{
\deduce{\Assert{x\in y\cup z}}{\Deny{x\in y}}
}
$
&
$
\infer[\AssertRule{\cup}\Elim{2}]{\Assert{x\in y}}{
\deduce{\Assert{x\in y\cup z}}{\Deny{x\in z}}
}
$
&
$
\infer[\DenyRule{\cup}\Elim{1}]{\Deny{x\in y}}{\Deny{x\in y\cup z}}
$
&
$
\infer[\DenyRule{\cup}\Elim{2}]{\Deny{x\in z}}{\Deny{x\in y\cup z}}
$
\end{tabular}
}
\end{center}

\begin{center}
\scalebox{0.85}{
\begin{tabular}{cccc}
$
\infer[\AssertRule{\cap}\Elim{1}]{\Assert{x\in y}}{\Assert{x\in y\cap z}}
$
&
$
\infer[\AssertRule{\cap}\Elim{2}]{\Assert{x\in z}}{\Assert{x\in y\cap z}}
$
&
$
\infer[\DenyRule{\cap}\Elim{1}]{\Deny{x\in z}}{
\deduce{\Deny{x\in y\cap z}}{\Assert{x\in y}}
}
$
&
$
\infer[\DenyRule{\cap}\Elim{2}]{\Deny{x\in y}}{
\deduce{\Deny{x\in y\cap z}}{\Assert{x\in z}}
}
$
\end{tabular}
}
\end{center}

\begin{center}
\scalebox{0.85}{
\begin{tabular}{cccc}
$
\infer[\AssertRule{\setminus}\Elim{1}]{\Assert{x\in y}}{\Assert{x\in y\setminus z}}
$
&
$
\infer[\AssertRule{\setminus}\Elim{2}]{\Deny{x\in z}}{\Assert{x\in y\setminus z}}
$
&
$
\infer[\DenyRule{\setminus}\Elim{1}]{\Assert{x\in z}}{
\deduce{\Deny{x\in y\setminus z}}{\Assert{x\in y}}
}
$
&
$
\infer[\DenyRule{\setminus}\Elim{2}]{\Deny{x\in y}}{
\deduce{\Deny{x\in y\setminus z}}{\Deny{x\in z}}
}
$
\end{tabular}
}
\end{center}

\begin{center}
\scalebox{0.85}{
\begin{tabular}{cccc}
$
\infer[\AssertRule{\times}\Elim{1}]{\Assert{x\in f\!st(y)}}{\Assert{x\in y\times z}}
$
&
$
\infer[\AssertRule{\times}\Elim{2}]{\Assert{x\in snd(z)}}{\Assert{x\in y\times z}}
$
&
$
\infer[\DenyRule{\times}\Elim{1}]{\Deny{x\in snd(z)}}{
\deduce{\Deny{x\in y\times z}}{\Assert{x\in f\!st(y)}}
}
$
&
$
\infer[\DenyRule{\times}\Elim{2}]{\Deny{x\in f\!st(y)}}{
\deduce{\Deny{x\in y\times z}}{\Assert{x\in snd(z)}}
}
$
\end{tabular}
}
\end{center}

\begin{center}
\scalebox{0.85}{
\begin{tabular}{cccc}
$
\infer[\AssertRule{\symdiff}\Elim{1a}]{\Assert{x\in z}}{
\deduce{\Assert{x\in y\symdiff z}}{\Deny{x\in y}}
}
$
&
$
\infer[\AssertRule{\symdiff}\Elim{1b}]{\Assert{x\in y}}{
\deduce{\Assert{x\in y\symdiff z}}{\Deny{x\in z}}
}
$
&
$
\infer[\AssertRule{\symdiff}\Elim{2a}]{\Deny{x\in z}}{
\deduce{\Assert{x\in y\symdiff z}}{\Assert{x\in y}}
}
$
&
$
\infer[\AssertRule{\symdiff}\Elim{2b}]{\Deny{x\in y}}{
\deduce{\Assert{x\in y\symdiff z}}{\Assert{x\in z}}
}
$
\end{tabular}
}
\end{center}

\begin{center}
\scalebox{0.85}{
\begin{tabular}{cccc}
$
\infer[\DenyRule{\symdiff}\Elim{1a}]{\Deny{x\in z}}{
\deduce{\Deny{x\in y\symdiff z}}{\Deny{x\in y}}
}
$
&
$
\infer[\DenyRule{\symdiff}\Elim{1b}]{\Deny{x\in y}}{
\deduce{\Deny{x\in y\symdiff z}}{\Deny{x\in z}}
}
$
&
$
\infer[\DenyRule{\symdiff}\Elim{2a}]{\Assert{x\in z}}{
\deduce{\Deny{x\in y\symdiff z}}{\Assert{x\in y}}
}
$
&
$
\infer[\DenyRule{\symdiff}\Elim{2b}]{\Assert{x\in y}}{
\deduce{\Deny{x\in y\symdiff z}}{\Assert{x\in z}}
}
$
\end{tabular}
}
\end{center}

\begin{center}
\scalebox{0.85}{
\begin{tabular}{cccc}
$
\infer[\AssertRule{}\subseteq\Elim{1}]{\Assert{z\in y}}{
\deduce{\Assert{x\subseteq y}}{\Assert{z\in x}}
}
$
&
$
\infer[\AssertRule{}\subseteq\Elim{2}]{\Deny{z\in x}}{
\deduce{\Assert{x\subseteq y}}{\Deny{z\in y}}
}
$
&
$
\infer[\DenyRule{\subseteq}\Elim{1}]{\Assert{f(x,y)\in x}}{
\Deny{x\subseteq y}
}
$
&
$
\infer[\DenyRule{\subseteq}\Elim{2}]{\Deny{f(x,y)\in y}}{
\Deny{x\subseteq y}
}
$
\end{tabular}
}
\end{center}

\begin{center}
\scalebox{0.85}{
\begin{tabular}{cccc}
$
\infer[\AssertRule{\exclusion}\Elim{1}]{\Deny{z\in y}}{
\deduce{\Assert{x\exclusion y}}{\Assert{z\in x}}
}
$
&
$
\infer[\AssertRule{\exclusion}\Elim{2}]{\Deny{z\in x}}{
\deduce{\Assert{x\exclusion y}}{\Assert{z\in y}}
}
$
&
$
\infer[\DenyRule{\exclusion}\Elim{1}]{\Assert{g(x,y)\in x}}{
\Deny{x\exclusion y}
}
$
&
$
\infer[\DenyRule{\exclusion}\Elim{2}]{\Assert{g(x,y)\in y}}{
\Deny{x\exclusion y}
}
$
\end{tabular}
}
\end{center}

\caption{The set of theory-specific rules extracted from the theory of sets.}
\label{fig:ts_sets_rules}
\end{figure}

There may be cases in which no theory-specific rule can be extracted from a given definitional axiom. For an example, consider the definitional axiom $(\forall x, y, z)(x\in y\cup z\imp x\in z\cup y)$. The two formulas in RINF equisatisfiable with this axiom are $x\in y\cup z\imp x\in z\cup y$ and $\lnot x\in z\cup y\imp \lnot x\in y\cup z$. In both cases, the correspondent rules violate the third theory-specific analytic restriction. Nevertheless, for each of the definitional axioms for the theory of sets we present here, it is possible to extract at least one theory-specific rule.

Now, we introduce an alternative version of cut for theory-specific proofs. To this end, we first provide two auxiliary definitions. For the first definition, let $r$ be a theory-specific rule with at least two premises. A premise $prem$ of $r$ is said to be a \textit{main premise} of $r$ if it contains all the variables occurring in the premises of~$r$. For example, $\Deny{x\in y\cap z}$ is a main premise of the rules $\DenyRule{\Elim{1}}$ and $\DenyRule{\Elim{2}}$ from $TS\_{sets}$. 
Although no rule in $TS\_{sets}$ has more than one main premise, this situation may in principle arise in other theories.

For the second definition, let $sf_1$ and $sf_2$ be two signed formulas and $ER$ be a set of expansion rules. We say that $sf_1$ and $sf_2$ are \textit{co-premises} in $ER$ if there exists a rule $r\in ER$ with at least two premises such that $sf_1$ and $sf_2$ can instantiate two different premises of~$r$. If $sf_1$ is a main premise of $r$, then $sf_2$ is called a \textit{minor co-premise} of $sf_1$ and $sf_1$ a \textit{major co-premise} of $sf_2$. For example, in the system $TS_{sets}$, the signed formula $\Deny{x\in y\cap(w\cup z)}$ is a major co-premise of $\Assert{x\in w\cup z}$, and $\Deny{x\in w}$ is a minor co-premise of $\Assert{x\in w\cup z}$.

We can now define a family of \textit{cut rules} for a set of theory-specific rules $TS$, as follows: 

\begin{center}
\begin{tabular}{p{2.5cm}}
$
\infer[cut]{\Assert{\varphi}\;\big|\Deny{\varphi}}{}
$
\end{tabular}

Either $\Assert{\varphi}$ or $\Deny{\varphi}$ is a minor co-premise\\in $TS$ of a node above
\end{center}

Observe that, by the first theory-specific analytic restriction, each theory-specific rule admits only finitely many instances. Hence, for any set of theory-specific formulas $ER$ and any $ER$-tableau $\texttt{T}$, the set of signed formulas in $\texttt{T}$ that can instantiate a main premise of a rule $r$ in $ER$ is also finite. Since the cut formula is required to be a minor co-premise within the set of theory-specific rules, it follows that the set of admissible cut instances is finite as well.

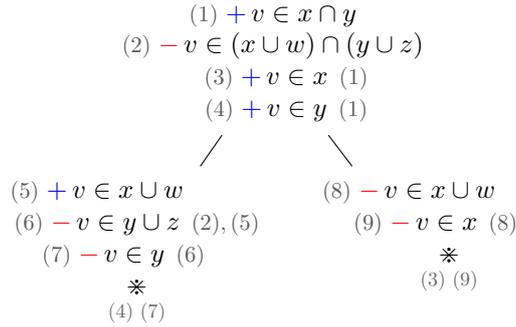
\begin{figure}[ht]
\centering
\begin{forest}
[$\hypnode{1}{\Assert{v\in x\cap y}}$ \\
    $\hypnode{2}{\Deny{v\in (x\cup w)\cap (y\cup z)}}$ \\
        $\quad\node{3}{\Assert{v\in x}}{(1)}$ \\
            $\quad\node{4}{\Assert{v\in y}}{(1)}$, align=center
            [$\hypnode{5}{\Assert{v\in x\cup w}}\quad\qquad$ \\
                $\node{6}{\Deny{v\in y\cup z}}{(2), (5)}$ \\
                    $\node{7}{\Deny{v\in y}}{(6)}\quad$ \\
                        $\absurd$ \\ $\close{(4) (7)}$, align=center]
            [$\hypnode{8}{\Deny{v\in x\cup w}}$ \\
                $\qquad\node{9}{\Deny{v\in x}}{(8)}$ \\
                        $\quad\qquad\absurd$ \\ $\quad\qquad\close{(3) (9)}$, align=center]
]
\end{forest}
\caption{A $TS_{sets}$-proof for $\set{\Assert{v\in x\cap y}, \Deny{v\in (x\cup w)\cap (y\cup z)}}$.}
\label{fig:tsproof_with_cut}
\end{figure}

In the $TS_{sets}$ proof of Figure \ref{fig:tsproof_with_cut}, we have an application of a cut to derive the nodes $(5)$ and $(8)$.  This particular cut instance is available in $TS_{sets}$, as $\Assert{v\in x\cup w}$ qualifies as a minor co-premise of $\Deny{v\in (x\cup w)\cap (y\cup z)}$ under the extraction procedure for the theory-specific rules.

\section{Proving complexity of proof exercises}\label{sec:complexity_proof_exercises}

The classification of an exercise by complexity should be grounded in the effort required to solve it.
Accordingly, if we are to claim that two proof exercises exhibit comparable levels of proving complexity, we must provide a principled way of determining how complex it is to prove them. In this section, we first explain why the theory-specific proofs presented in Section \ref{sec:Tsproofs} constitute a suitable alternative to directly translating informal proofs into formal ones. 
We then introduce formal definitions of proving complexity for theory-specific proofs, enabling us to determine when two refutable sets of signed formulas have a comparable level of proving complexity. We will use this notion of proving complexity to control the difficulty level of the exercises generated by the procedure described in Section~\ref{sec:main_procedure_description}.

\subsection{Formalizing informal proofs}

Novice students in higher Mathematics are typically trained to prove theorems through informal proofs. They learn how to manipulate the givens and the goals of a theorem by means of proof strategies expressed in sentences that combine natural and mathematical language. These sentences are then arranged into a continuous text ---an informal proof--- whose structure, although logically coherent, lacks an explicit formal organization.  
In fact, as Lamport has argued in \cite{Lamport:1995}, adherence to the narrative prose style common in nineteenth-century mathematics often obscures the logical structure of proofs and conceals the dependency relations essential to their verification.

If we aim to determine when two proof exercises have a comparable level of proving complexity, an important question is: How do we devise a way of precisely measuring the proving complexity of such unstructured objects? In what follows, we argue that the theory-specific proofs introduced in Section~\ref{sec:Tsproofs} may be regarded as a formalization of informal proofs. By exploiting the tree structure inherent in formal proofs, it becomes possible to define a method for calculating proving complexity, as detailed in Subsection~\ref{subsec:comparability_complexity}.

Students are typically introduced to mathematical proofs without prior training in formal logic. As a result, the informal proofs they learn to write often leave implicit some of the logical steps underlying the proof strategies they employ. This observation plays a central role in our proposal and underpins our adoption of theory-specific proofs as a formal counterpart to informal proofs. Consider, for instance, the ``element argument'', illustrated in Figure~\ref{fig:element_argument}, which has been described as ``the fundamental proof technique of set theory'' \cite{Epp2010}:

\begin{figure}[h]
\centering
\includegraphics[width=0.8\textwidth]{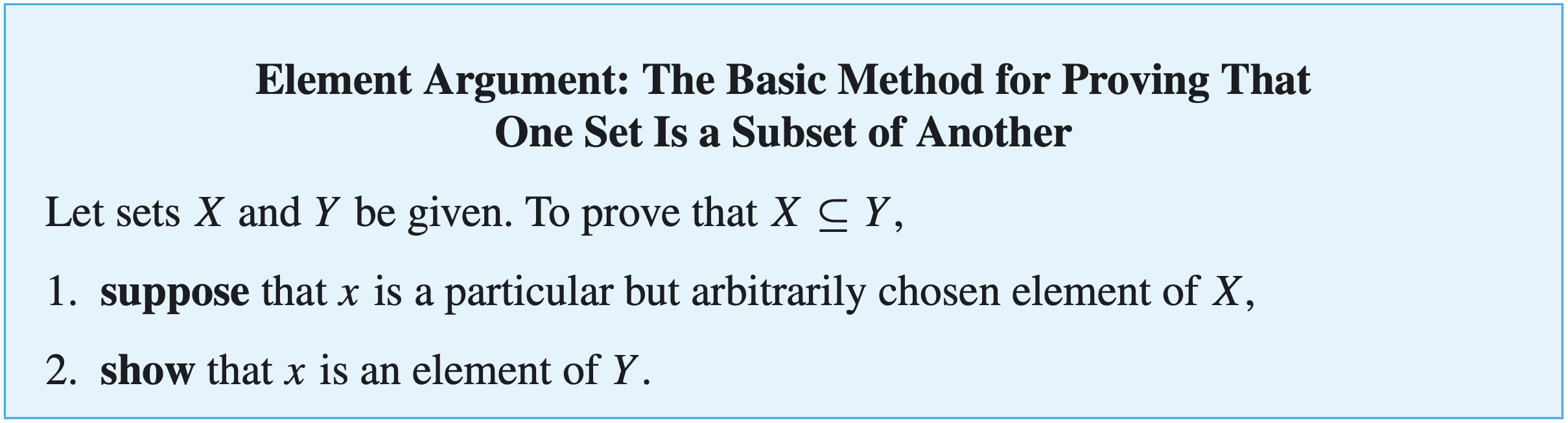}
\caption{Description of the “Element Argument” from \cite{Epp2010}.}
\label{fig:element_argument}
\end{figure}

This proof strategy is typically employed without making explicit what goes behind the scenes: in logical terms, inclusion is defined in terms of a quantified formula in prenex normal form whose matrix is an implication between assertions involving the membership relation.

\begin{figure}[h]
\centering
\includegraphics[width=0.8\textwidth]{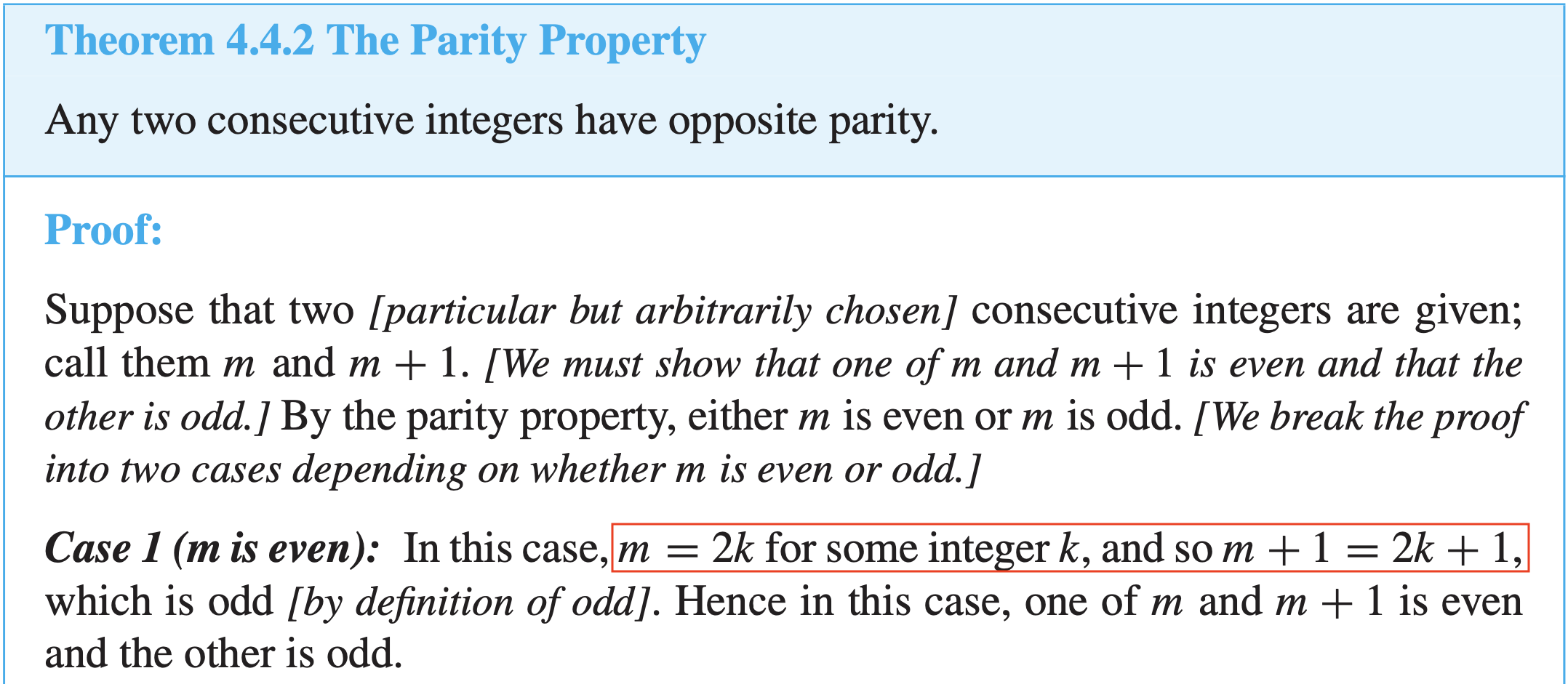}
\caption{A fragment of the proof from \cite{Epp2010} that any two consecutive integers have opposite parity.}
\label{fig:epp_parity}
\end{figure}

A further illustration of how logical structure remains implicit in informal proofs can be found in the excerpt shown in in Figure \ref{fig:epp_parity} of a proof that two consecutive integers have opposite parity. From a formal perspective, the step from $m = 2k$ to $m + 1 = 2k + 1$ relies on the substitutivity of identicals, more precisely on equality being a congruence relation with respect to which addition (or actually successor, in this case) is compatible. However, no justification for this step is made explicit in the proof.

\begin{figure}[h]
\centering
\includegraphics[width=1.0\textwidth]{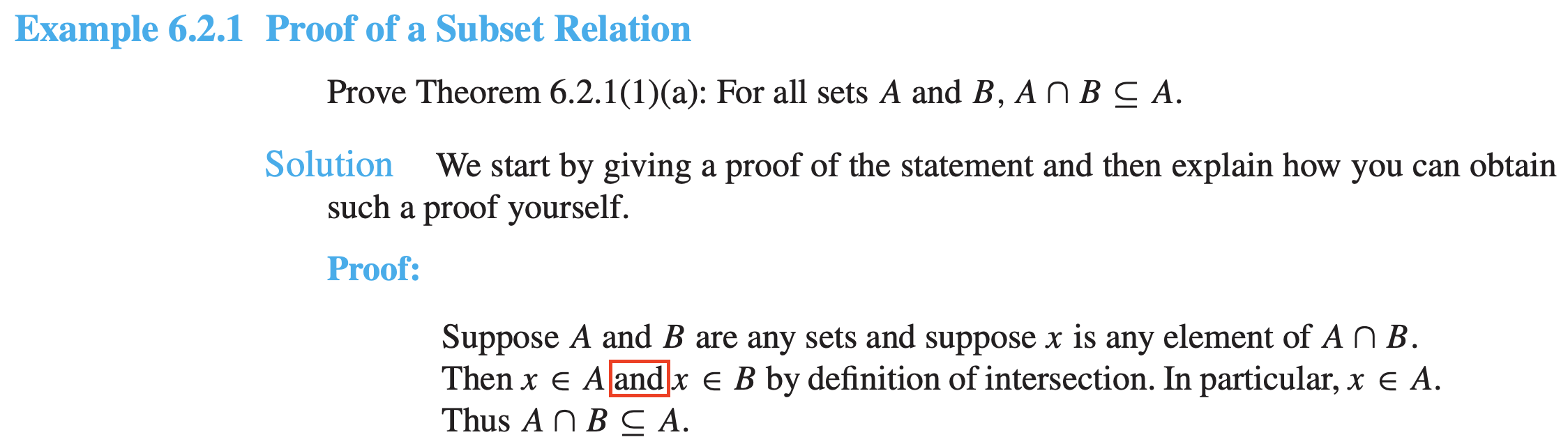}
\caption{A proof that that $A \cap B \subseteq A$ for all sets $A$ and $B$, adapted from \cite{Epp2010}.}
\label{fig:epp_sets}
\end{figure}

The fact that logical steps are oftentimes left implicit in informal proofs does not mean that they are always concealed. Consider, for instance, the example shown in Figure \ref{fig:epp_sets} of a proof that $A \cap B \subseteq A$ for all sets~$A$ and~$B$. Do note that the role of the conjunction in the definition of intersection is played by the particle ``and'' highlighted in the red rectangle. This reflects the usual definition of intersection, which claims that ``$x \in X \cap Y$ if and only if $x \in X$ and $x \in Y$”. Now consider, however, an alternative definition of intersection presented by the following three unidirectional statements: 
\begin{enumerate}
\item If $x \in X \cap Y$, then $x \in X$.
\item If $x \in X \cap Y$, then $x \in Y$.
\item If $x \in X$, then if $x \in Y$, then $x \in X \cap Y$.
\end{enumerate}

In this example, one could also directly conclude that $x \in A$ from the fact that $x \in A \cap B$ without making explicit the ``and'' corresponding to the conjunction.

\subsection{Comparability of proving complexity}\label{subsec:comparability_complexity}

If we adopt theory-specific proofs as a formal counterpart of informal proofs, what would be the logical description of the statement of a proof exercise?  
Suppose $E$ is a proof exercise. We can say that the formulas of the givens of~$E$ form a set $\Gamma:= \set{\gamma_1, \dots, \gamma_m}$ and the formulas of the goal of~$E$ form a set $\Delta := \set{\delta_1, \dots, \delta_n}$. Then, the consequence statement $\Gamma\vdash\Delta$ and, hence, the refutable set $\{\Assert\gamma_1, \dots, \Assert\gamma_m,$ $\Deny\delta_1, \dots, \Deny\delta_n\}$ could be the logical description of~$E$. Do note, however, that the set of formulas that logically describes $E$ may contain propositional connectives and cannot be converted into an equisatisfiable set into STSNF. Such exercises will not fall under the scope of our method. In this subsection, aiming to be able to state when two proof exercises have a comparable level of proving complexity, we provide a notion of comparability of proving complexity between refutable sets of theory-specific signed formulas.

Let $ER$ be a set of theory-specific rules for a definitional theory $Th$ and let $SF_1$ and $SF_2$ be two sets of theory-specific signed formulas of $L$ that are $ER$-refutable. Intuitively, if the minimal effort required to prove $SF_1$ and $SF_2$ yields $ER$-proofs with the same structure, then we want to regard these two sets as having comparable levels of proving complexity. We start by defining the notion of isomorphism between proofs. To do so, we resort to the tree structure that tableau proofs have so we can rely on the concept of isomorphism between trees.

A precise definition of isomorphism between tableau proofs that adequately captures their deductive comparability cannot rely solely on the tree structure of the tableaux involved. This becomes apparent when comparing the tableau proofs displayed in Figure~\ref{fig:non_deductive_simlarity}, both constructed using the set of rules $TS_{sets}$. Although each tableau contains~$6$ nodes, they were constructed in slightly different ways: in (a), a two-premise rule is applied once, whereas in (b) no such rule is used. Moreover, tableau (a) is clean, while tableau (b) is not.
For this reason, the formal notion of deductive isomorphism we introduce below is grounded in two complementary forms of isomorphism, defined below: one at the level of formulas and another at the level of what we call ``justification trees''.

\begin{figure}[h]
\centering
\begin{tabular}{cc}
\begin{tabular}[t]{c}
\hspace{.2cm}(a) 
\vspace{0.1cm}\\
\begin{forest}
[$\hypnode{1}{\Assert{p_1\in\complement p_3\setminus p_2}}\quad$ \\
    $\hypnode{2}{\Assert{p_1\in p_2\cup p_3}}\quad$ \\ 
        $\node{3}{\Assert{p_1\in\complement p_3}}{(1)}$ \\
            $\node{4}{\Deny{p_1\in p_2}}{(1)}$ \\
                $\node{5}{\Deny{p_1\in p_3}}{(3)}$ \\
                    $\quad\;\,\node{6}{\Assert{p_1\in p_3}}{(2), (4)}$\\ $\absurd$ \\ $\close{(5) (6)}$, align=center
]
\end{forest} \\
\end{tabular}
&
\begin{tabular}[t]{c}
(b)\hspace{.0cm} 
\vspace{0.1cm}\\
\begin{forest}
[$\hypnode{1}{\Assert{p_1\in p_2\cap(p_3\cap(p_4\cap p_5))}}\quad\;\;$ \\
    $\hypnode{2}{\Deny{p_1\in p_5}}\quad\;\;$ \\ 
        $\node{3}{\Assert{p_1\in p_2}}{(1)}$ \\
            $\node{4}{\Assert{p_1\in p_3\cap(p_4\cap p_5)}}{(1)}$ \\
                $\node{5}{\Assert{p_1\in p_4\cap p_5}}{(4)}$ \\
                    $\,\node{6}{\Assert{p_1\in p_5}}{(5)}$\\ $\absurd$ \\ $\close{(2) (6)}$, align=center
]
\end{forest}\\
\end{tabular}
\\
\end{tabular}
\caption{Tableau proofs for (a) $\set{\Assert{p_1 \in \complement p_3 \setminus p_2}, \Assert{p_1 \in p_2 \cup p_3}}$ and $(b)$ $\{\Assert{}p_1 \in p_2 $ $\cap\, (p_3 \cap (p_4 \cap p_5)), $ $\Deny{}p_1 \in p_5\}$.}
\label{fig:non_deductive_simlarity}
\end{figure}

Given a language $L$, two theory-specific formulas $\varphi_1$ and $\varphi_2$ of $L$ are \textit{syntactically isomorphic} if the abstract representation of $\varphi_1$ is isomorphic to the abstract representation of $\varphi_2$ and the set of variables of $\varphi_1$ is equal to the set of variables of $\varphi_2$. We extend this definition to signed formulas by saying that two signed formulas $sf_1$ and $sf_2$ are syntactically isomorphic if $fmla(sf_1)$ and $fmla(sf_2)$ are syntactically isomorphic.

Consider the following three formulas from the theory of sets: $\varphi_1:= x\subseteq y\cap z$, $\varphi_2:= z\in x\cap y$ and $\varphi_3 := x\subseteq x\cup y$. The formulas $\varphi_1$ and $\varphi_2$ are syntactically isomorphic, but they are not syntactically isomorphic to $\varphi_3$, since the variable $z$ is present in both $\varphi_1$ and $\varphi_2$, but not in $\varphi_3$. If we did not impose the equality between the sets of variables, a theory-specific formula could have an infinite amount of syntactically isomorphic formulas. Hence, this would not allow us to implement a procedure that searches for all syntactically isomorphic formulas to a given formula as we do in Section \ref{sec:main_procedure_description}.

\begin{figure}[h]
\centering
\begin{tabular}{cc}
\multicolumn{2}{c} {
(a)
}\\
\begin{tabular}[t]{c}
\begin{forest}
[$\hypnode{1}{\Assert{x\in y\cap z}}\;\;\quad$ \\
    $\hypnode{2}{\Deny{x\in y}}\;\;\quad$ \\ 
        $\node{3}{\Assert{x\in y}}{(1)}$ \\ $\absurd$ \\ $\close{(2) (3)}$, align=center
]
\end{forest}\\
\end{tabular}
&
\begin{tabular}[t]{c}
\begin{forest}
[$\absurd$
    [$\hypnode{2}{\Deny{x\in y}}$]
    [$\hypnode{3}{\Assert{x\in y}}$
        [$\hypnode{1}{\Assert{x\in y\cap z}}$]
    ]
]
\end{forest}\\
\end{tabular}
\\
\multicolumn{2}{c} {
(b)
}\\
\begin{tabular}[t]{c}
\begin{forest}
[$\hypnode{1}{\Assert{x\in y\setminus z}}\;\;\quad$ \\
    $\hypnode{2}{\Deny{x\in y}}\;\;\quad$ \\ 
        $\node{3}{\Assert{x\in y}}{(1)}$ \\ $\absurd$ \\ $\close{(2) (3)}$, align=center
]
\end{forest}\\
\end{tabular}
&
\begin{tabular}[t]{c}
\begin{forest}
[$\absurd$
    [$\hypnode{2}{\Deny{x\in y}}$]
    [$\hypnode{3}{\Assert{x\in y}}$
        [$\hypnode{1}{\Assert{x\in y\setminus z}}$]
    ]
]
\end{forest}\\
\end{tabular}
\end{tabular}
\caption{(a) A proof for $\set{\Assert{x \in y \cap z}, \Deny {x \in y}}$ and its justification tree and (b) a proof for $\set{\Assert{x \in y \setminus z}, \Deny{x \in y}}$ and its justification tree.}
\label{fig:deduc_iso_sf1}
\end{figure}
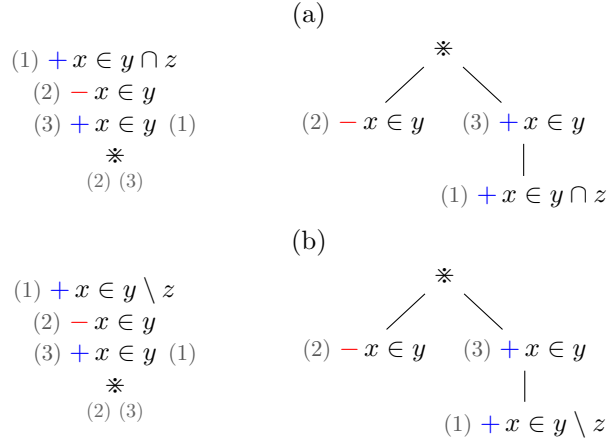

Let $ER$ be a set of theory-specific rules for some definitional theory. For a branch $\theta$ of an $ER$-proof $\texttt{P}$, a \textit{justification tree} $J$ for $\theta$ has as root a node whose content is the symbol $\closure$ and the other nodes of $J$ are nodes of $\theta$. Moreover, the children of the root node are the nodes that made the branch $\theta$ close in $\texttt{P}$. For any other node $i$ in $J$, its children in $J$ are the nodes that justify the addition of $i$ to~$\theta$. Given an $ER$-proof $\texttt{P}$ containing branches $\theta_1, \dots, \theta_n$, if $J_1, \dots, J_n$ are, respectively, the justification trees for $\theta_1, \dots, \theta_n$, we can also say that $J_1, \dots, J_n$ are the justification trees for $\texttt{P}$. In Figure \ref{fig:deduc_iso_sf1}, we present a proof for $\set{\Assert{x \in y \cap z}, \Deny{x \in y}}$, a proof for $\{\Assert{x}$ $\in y \setminus z, \Deny{x \in y}\}$ and their respective justification trees. 

Let $\theta_R$ and $\theta_L$ be two branches of theory-specific tableau proofs. Given justification trees $R$ for $\theta_R$ and $L$ for $\theta_L$, we say that $R$ and $L$ are \textit{deductively isomorphic} when there is an isomorphism between $R$ and $L$ that preserves the syntactic isomorphism of their formulas. We extend this notion to proofs by saying that two theory-specific proofs $\texttt{P}_1$ and $\texttt{P}_2$ are \textit{deductively isomorphic} if there is a one-to-one correspondence between the branches of $\texttt{P}_1$ and $\texttt{P}_2$ mapping branches of the same length having deductively isomorphic justification trees. In Figure \ref{fig:deduc_iso_sf1}, the two proofs are deductively isomorphic. 

The deductive isomorphism of proof relies on the deductive isomorphism of their justification trees because there can be isomorphic tableaux with different proof structures. For example, a tableau proof for $\set{\Assert{x\in y\setminus y}}$ could be constructed with the application of rules $\AssertRule{\setminus}\Elim{1}$ and $\AssertRule{\setminus}\Elim{2}$. This tableau would be isomorphic to the ones presented in Figure \ref{fig:deduc_iso_sf1}, but the latter tableaux were constructed with only one rule application.

Let $ER$ be a set of theory-specific expansion rules for some definitional theory. The \textit{deductive size} of a branch $\theta$ of an $ER$-proof is given by the amount of nodes of its justification tree and the \textit{deductive size} of an $ER$-proof is given by the sum of the sizes of its branches. The deductive size of the proofs in Figure \ref{fig:deduc_iso_sf1} is~4 and the deductive size of the proof in Figure \ref{fig:tsproof_with_cut} is 13.

Let $ER$ be a set of theory-specific expansion rules for a definitional theory $\truple{L}{Ax}{<}$. Given a set of theory-specific signed formulas $SF$ of $L$, a clean $ER$-proof $\texttt{P}$ for $SF$ is \textit{minimal} if there is no other clean $ER$-proof for $SF$ whose deductive size is smaller than the deductive size of $\texttt{P}$. The set of minimal proofs for a set of signed formulas provides the basis for determining its \textit{proving complexity}.

We now have tools that allow us to formalize the notion of comparability of proving complexity between sets of signed formulas. Let $ER$ be a set of theory-specific expansion rules for a definitional theory $\truple{L}{Ax}{<}$. For any sets $SF_1$ and $SF_2$ of theory-specific signed formulas of $L$, we say that $SF_1$ and $SF_2$ have a \textit{comparable level of proving complexity} when there are: (i) an $ER$-proof $\texttt{P}_1$ for $SF_1$ that is minimal and there is an $ER$-proof $\texttt{P}_2$ for $SF_2$ that is also minimal such that $\texttt{P}_1$ and $\texttt{P}_2$ are deductively isomorphic and (ii) there is a bijective correspondence between $SF_1$ and $SF_2$ mapping syntactically isomorphic signed formulas with equal signs. In this case, $SF_1$ and $SF_2$ are also said to be \textit{proof-isomorphic}.

Recall that we use sets of signed formulas to represent a consequence relation that, in turn, formally describes the statement of a proof exercise. The clause (ii) in the previous definition is specified to guarantee that the structure of givens and goals of proof exercises with comparable levels of proving complexity is preserved. 

The proofs presented in Figure \ref{fig:deduc_iso_sf1} are minimal, since they are constructed with a single application of an expansion rule and there are no proofs with no applications of expansion rules for these sets of signed formulas. Moreover, the associated justification trees are deductively isomorphic. Thus, $\set{\Assert{x \in y \cap z}, \Deny{x \in y}}$ and $\set{\Assert{x \in y \setminus z}, \Deny{x \in y}}$ have a comparable level of proving complexity.

\section{Procedure of generation of proof exercises with comparable levels of proving complexity}\label{sec:main_procedure_description}

Recall that we formally describe a proof exercise by a set of signed formulas in STSNF.
Given a definitional theory $Th := \truple{L}{Ax}{<}$, the procedure we provide for the automated generation of proof exercises with comparable levels of proving complexity has as inputs a set of signed formulas (the logical description of the exercise) of $L$ and a set of theory-specific rules for $Th$. We, then, divide the procedure into two main steps:
\begin{enumerate}
\item Search for minimal proofs which provide the basis for determining the proving complexity of the exercise given as input.
\item Search for proof-isomorphic sets of signed formulas. Each of these sets of signed formulas represents the exercises with a level of proving complexity comparable to that of the proof exercise whose logical description was given as input.
\end{enumerate}

It is worth mentioning that what our method generates are not exactly questions, but mathematical objects that lecturers may use to design questions for their students. This does not prevent such method to be considered one of AQG. Let $c$ be an output of the tool. We may assume that what the tool actually generates are questions in the format of ``What could be a proof for $c$?''. In addition, the knowledge source may be seen as the set of theory-specific rules which we use to extract the proofs, while the desired property of the output, that is its level of proving complexity, is encapsulated in the input exercise.

In the following subsections, we describe the two main steps of the procedure of generation. A prototype implementation of the whole procedure was developed and is available at \url{https://github.com/joaomendesln/GePECC}.

\subsection{Search for minimal proofs}\label{subsec:min_proofs}

Before providing the procedure, we introduce an additional auxiliary definition. Let $ER$ be a set of theory-specific rules and $SF$ a set of signed formulas. We define the set $T_n(ER, SF)$ as the set of all $ER$-tableaux for $SF$ with~$n$ applications of rules of $ER$, including the cut for $ER$. Since both the rules of $ER$ and the cut for $ER$ have a finite amount of instances, then the set $T_n(ER, SF)$ is finite.

Given a definitional theory $Th := \truple{L}{Ax}{<}$, a set $ER$ of theory-specific rules for $Th$ and an $ER$-refutable set of signed formulas $SF$ of $L$, the procedure outputs the set of all minimal $ER$-proofs for $SF$. It starts by constructing the set $T_1(ER, SF)$, followed by the construction of the set $T_2(ER, SF)$ and so on. For each $T_n(ER, SF)$ set, we check if each tableau in $T_n(ER, SF)$ is an $ER$-proof for $SF$. When the first $ER$-proof $\texttt{P}$ for $SF$ is found, the deductive size of $\texttt{P}$ is set as the upper bound and $\texttt{P}$ is added to the output set $M$ of minimal proofs. The construction of the sets $T_i(ER,SF)$ continues and the set of minimal proofs is updated whenever an $ER$-proof $\texttt{S}$ for $SF$ is found. If the deductive size of $\texttt{S}$ is equal to the current upper bound, then $\texttt{S}$ is added to $M$. If the deductive size of $\texttt{S}$ is smaller than the current upper bound, then $M$ is emptied, $\texttt{S}$ is added to it and the new upper bound is the deductive size of~$\texttt{S}$. The procedure terminates after the construction of the set $T_{up}(ER,SF)$, where $up$ is the current upper bound.

Regarding the termination of the procedure, we sequentially check the existence of proofs in all the elements of the $T_n(ER,SF)$ sets. As these sets are always finite, their construction terminates in a finite amount of time and, hence, such step is not an issue. 
Furthermore, the procedure is guaranteed to eventually set an upper bound during the search for minimal proofs, since a precondition of the procedure is that the input set of signed formulas $SF$ is $ER$-refutable.

One threat for the correctness of this procedure is the existence of a minimal proof that is not added to the output set, that is, for a set of signed formulas $SF$ and a set of theory-specific rules $ER$, there is a proof $\texttt{P}$ of size $m$ in the set $T_n(ER,SF)$, where $n>m$ and $m$ is the size of the proofs in the output set. This can happen if some nodes of $\texttt{P}$ are not part of any of the justification trees for $\texttt{P}$. However, in this case, the proof $\texttt{P}$ is not clean and, by definition, $\texttt{P}$ cannot be a minimal proof.

\subsection{Search for proof-isomorphic sets of signed formulas}\label{subsec:proof_iso_procedure}

Given a minimal $ER$-proof $\texttt{P}$ extracted for a set of signed formulas $SF$ of a language $L$, we provide a procedure that searches for all sets of signed formulas of $L$ having a minimal proof that is deductively isomorphic to $\texttt{P}$. To exemplify a simple, but not necessarily efficient, approach to this, consider the conjecture $\{\Assert{x\in y\cap(w\cup z)}, \Deny{x}\in (y\cap w)$ $\cup\, z\}$ and the minimal proof $\texttt{M}$ for it described in Figure \ref{fig:search_proof_iso_min}. For every set $\{\Assert{}x\,\square_1\, y\circ_1(w $ $\circ_2\;  z),$ $\Deny{x\,\square_2\, (y\circ_1 w)\circ_2 z}\}$ such that
$\square_1$ and $\square_2$ range among the predicate symbols in $\set{\in,\subseteq,\exclusion}$ and $\circ_1$, $\circ_2$, $\circ_3$ and $\circ_4$ range among the function symbols in $\set{\cup,\cap,\setminus,\symdiff,\times}$,
we could try to construct a proof that is deductively isomorphic to $\texttt{M}$. It would not make sense to check, however, if there is a minimal proof for the set $SF_1:=\{\Assert{}x\in y\symdiff(w\,\cup$ $z), \Deny{x}$ $\subseteq (y\cap w)\cup z\}$ that is deductively isomorphic to $\texttt{M}$. First, after an application of the rule $\DenyRule{\subseteq}\Elim{}$ having as premise $\Deny{}x\subseteq (y\cap w)\,\cup\, z$, the skolem function symbol~$f$ will appear in the proof and there is no skolem function symbol in~$\texttt{M}$. Second, all the rules involving $\symdiff$ are $2$-premise rules. After all, for this small example, there would it be $5625$ cases to test including sets of signed formulas such as $SF_1$.

\begin{figure}[h]
\centering

\begin{forest}
[$\hypnode{1}{\Assert{x\in y\cap(w\cup z)}}$ \\
    $\hypnode{2}{\Deny{x\in (y\cap w)\cup z}}$ \\ 
        $\quad\node{3}{\Assert{x\in y}}{(1)}$ \\ 
            $\quad\node{4}{\Assert{x\in w\cup z}}{(1)}$ \\ 
                $\quad\node{5}{\Deny{x\in y\cap w}}{(2)}$ \\ 
                    $\quad\node{6}{\Deny{x\in z}}{(2)}$ \\
                        $\quad\quad\;\,\node{7}{\Deny{x\in w}}{(3), (5)}$\\
                            $\quad\quad\;\,\node{8}{\Assert{x\in w}}{(4), (6)}$\\ $\quad\quad\absurd$ \\ $\quad\quad\close{(7) (8)}$, align=center
]
\end{forest}
\vspace{-1em}
\caption{A minimal proof $M$ for $\set{\Assert{x\in y\cap(w\cup z)}, \Deny{x\in (y\cap w)\cup z}}$.}
\label{fig:search_proof_iso_min}
\end{figure}

Before presenting the procedure, we point out how we can restrict the search space of proof-isomorphic candidate signed formulas. 
Informally, given a minimal $ER$-proof $\texttt{P}$ extracted for a set of signed formulas $SF$ over a language $L$, we generate candidate substitutions of a symbol occurring in a signed formula of $SF$ by other symbols of~$L$. The structure of~$\texttt{P}$ and the rules of $ER$ constrain this process, determining which substitutions are admissible.
We formalize such notion in what follows.

The \textit{occurrence $n$ of a symbol $s$ on a formula $\varphi$} is the $n$-th occurrence of $s$ in $\varphi$ considering its concrete representation, that is, the sequence resulting from the in-order traversal in the abstract representation of $\varphi$. In a theory-specific formula $\varphi$, a predicate symbol $p$ has at most one occurrence in $\varphi$. However, the same cannot be claimed about function symbols. This is why we introduce the following definition.

Considering the abstract representation of a theory-specific formula $\varphi$, every symbol of $\varphi$ may be represented by a \textit{position}, which is a string of natural numbers concatenated by the symbol ``$\cdot$''. Let $S$ be the abstract representation of $\varphi$. So, the position of the root of $S$ is represented by the empty string~$\epsilon$. For any other node $s$ of $S$, the position of $s$ is given by $i{\cdot}\pi$, where $s$ is the $i$-th child of the node $t$ and the position of $t$ is given by $\pi$. The \textit{position of the occurrence $o$ of a function symbol $f$ in a formula $\varphi$} is denoted by $pos(o,f,\varphi)$. For example, for the formula $\varphi_1:=x\in (y \cap w)\,\cap\, z$, we have that $pos(1,\cap,\varphi_1)=2$ and $pos(2,\cap,\varphi_1)=1{\cdot}2$. 

Let $s_1$ and $s_2$ be two symbols of a language $L$. A symbol $s_2$ is a \textit{deductive matching symbol} of the occurrence $o_1$ of $s_1$ in a premise $prem_1$ of a rule $r_1\in ER$ when there is a rule $r_2$ in $ER$ such that: (i) the premises of $r_1$ and the premises of $r_2$ are syntactically isomorphic, (ii) the conclusion of $r_1$ is isomorphic to the conclusion of $r_2$ and (iii) the symbol $s_2$ appears in a premise $prem_2$ of $r_2$, where $fmla(prem_1)$ and $fmla(prem_2)$ are isomorphic and $pos(o_2,s_2,fmla(prem_2))=pos(o_1,s_1,fmla(prem_1))$ for some occurrence $o_2$ of $s_2$ in $fmla(prem_2)$. When there is only one occurrence of $s_1$ in the premises of $r_1$, we simply say that $s_2$ is a deductive matching symbol of $s_1$ with respect to the rule $r_1$.

In the set of rules $TS_{sets}$, the set of deductive matching symbols of $\cap$ with respect to the rule $\AssertRule{\cap}\Elim{}$ is $\set{\cap, \cup, \setminus}$. But, in the same set of rules, the set of deductive matching symbols of $\cap$ with respect to the rule $\DenyRule{\cap}\Elim{1}$ is $\set{\cap, \cup, \setminus, \symdiff}$. 

Let $r$ be an $n$-premise theory-specific rule, whose premises are $prem_1,\dots,prem_n$. The rule $r$ is applicable to a tableau $T$ if there are nodes $m_1,\dots,m_n$ in $T$ that are instances of $prem_1,\dots,prem_n$ respectively. In this case, we say that the nodes $m_1,\dots,m_n$ justify the application of $r$ in $T$ and that $m_i$ is the \textit{justification node} of the premise $prem_i$ for this application of $r$ in $T$, where $1\leq i\leq n$. In the proof $\texttt{M}$, the nodes $(3)$ and $(5)$ justify the application of $\DenyRule{\cap}\Elim{1}$, and the node $(1)$ justifies the application of $\AssertRule{\cap}\Elim{}$ in such proof.

Let $ER$ be a set of theory-specific rules and $\texttt{T}$ be an $ER$-tableau. Now, let $r$ be a rule of $ER$ having one of its premises the signed formula $prem$, $n$ be a node of $T$ such that $n$ is the justification of $prem$ for an application of $r$ in $\texttt{T}$ and $s$ be a symbol that occurs in $prem$ and in $n$. We say that the occurrence $o_2$ of $s$ in $prem$ is the \textit{justification matching symbol occurrence} of the occurrence $o_1$ of $s$ in $n$ if $pos(o_2,s,fmla(prem)) = pos(o_1,p_1,fmla(n))$.

For example, in the proof $\texttt{M}$, the justification matching symbol of the only occurrence of $\cap$ in the node $(1)$ is its single occurrence in the premise $\Assert{x\in y\cap z}$. Then, in the procedure we aim to present, we can try to replace $\cap$ in $(1)$ by any of the symbols of the set $\set{\cap,\cup,\setminus}$, since this set represents the deductive matching symbols of the occurrence of $\cap$ in the premise of $\AssertRule{\cap}\Elim{}$. However, this is still not enough. The occurrence of $\cup$ in the node $(1)$, for example, has no matching justification symbols, since $\cup$ does not even appear in the premise of $\AssertRule{\cap}\Elim{}$. For the case of such occurrences, we introduce the next couple of definitions.

Let $m$ be a direct descendant node of $n$ in a proof $\texttt{P}$ and $r$ be the rule that was applied to $\texttt{P}$ such that is $n$ a justification node of a premise $prem$ for this application of $r$. The occurrence $o_1$ of a function symbol $f_1$ in $fmla(m)$ is a \textit{direct descendant occurrence} of the occurrence $o_2$ of a function $f_2$ in $fmla(n)$ if the symbol in the position $pos(o_1,f_1,fmla(m))$ of the formula in the conclusion of~$r$ and the symbol in the position $pos(o_2,f_2,fmla(n))$ of $fmla(prem)$ are the same variable. The set of \textit{descendant occurrences} of the occurrence~$o$ of a function symbol $f$ in $n$ is the reflexive-transitive closure of the direct descendant occurrences of the occurrence $o$ of $f$ in $\texttt{P}$.

In Figure \ref{fig:descendant_occurrences}, the occurrence 2 of $\cap$ in (3), highlighted in green, is a direct descendant of the occurrence 3 of $\cap$ in (1), highlighted in blue. The occurrence 1 of $\cap$ in (1) has no direct descendant occurrences, and the same holds for the occurrence of $\cap$ in (4), highlighted in red. The descendant occurrences of the occurrences of $\cap$ highlighted in blue are the occurrences of it highlighted in blue, green and red. Now, we can define the set of candidate signed formulas to which we restrict our search in this step of the generation.

\begin{figure}[h]
\centering

\begin{forest}
[$\hypnode{1}{\Assert{v\in x\cap (y\cap (w \highlight{CornflowerBlue}{\cap} z))}}$ \\
    $\hypnode{2}{\Deny{v\in z}}$ \\ 
        $\;\quad\node{3}{\Assert{v\in y\cap (w \highlight{Emerald}{\cap} z)}}{(1)}$ \\ 
            $\;\quad\node{4}{\Assert{v\in w \highlight{LightRed}{\cap} z}}{(3)}$ \\
                $\;\quad\node{5}{\Assert{v\in z}}{(4)}$ \\
    $\quad\quad\absurd$ \\ $\quad\quad\close{(2) (5)}$, align=center
]
\end{forest} 
\vspace{-1em}
\caption{A minimal proof for $\set{\Assert{v\in x\cap (y\cap (w \cap z))}, \Deny{v\in z}}$.}
\label{fig:descendant_occurrences}
\end{figure}

Let $SF$ be a set of signed formulas of $L$, $ER$ be a set of expansion rules, $\texttt{T}$ be an $ER$-proof for $SF$ and $sf\in SF$ be such that $sf=fmla(n)$ for some node in $n$ of $\texttt{T}$ that is a justification for an application of the rule $r$ of $ER$. If $p$ is a predicate symbol that occurs in $sf$, then the set of \textit{proof-isomorphic predicate symbols} of $p$ in $sf$ is either: (i) the set of deductive matching symbols of $p^\prime$ in $ER$ with respect to $r$ if $p^\prime$ is a matching justification of $p$ in $\texttt{T}$, or (ii) the set of predicate symbols of $L$ if $p$ does not have a matching justification symbol in $\texttt{T}$. If $f$ is a function symbol occurring in $sf$, the set of \textit{proof-isomorphic function symbols} of the occurrence $o$ of $f$ in $sf$ is either: (i) the set of deductive matching symbols of all matching justification symbols of descendant occurrences of $f$ in $\texttt{T}$, or (ii) the set of function symbols of $L$ if none of the descendant occurrences of $f$ in $\texttt{T}$ have matching justification symbols in $\texttt{T}$.

Let $SF$ be a set of signed formulas, $ER$ be a set of expansion rules and $\texttt{P}$ be an $ER$-proof for $SF$. A set of signed formulas $SF^\prime$ is a \textit{proof-isomorphic candidate set of signed formulas} for $SF$ with respect to $\texttt{P}$ in $ER$ if $SF^\prime = SF$ or $SF^\prime$ is the result of replacing at least one occurrence of predicate (or function) symbol $s$ of a signed formula $sf\in SF$ by a proof-isomorphic predicate (or function) symbol of such $s$ in $sf$.

Consider the set $\set{\Assert{x\in y\cap(w\cup z)}, \Deny{x\in (y\cap w)\cup z}}$ in the proof of Figure \ref{fig:search_proof_iso_min}. We calculate the size of the set of proof-isomorphic candidate signed formulas for them. For this, we refer to $\Assert{x\in y\cap(w\cup z)}$ as $sf_1$, and to $\Deny{x\in (y\cap w)\cup z}$ as $sf_2$. The occurrence of $\cap$ in $sf_1$ and the occurrence of $\cup$ in $sf_2$ may be replaced by either $\cap$, $\cup$ or $\setminus$. The occurrence of $\cap$ in $sf_1$ and the occurrence of $\cup$ in $sf_2$ may be replaced by either $\cap$, $\cup$, $\setminus$ or $\symdiff$. Then, the size of the set is $3 \times 3 \times 4 \times 4 = 144$ and this reduces, for this example, almost $40$ times the amount of cases to be tested.

Using the above definitions, we can now describe our procedure for searching for proof-isomorphic sets of signed formulas. As input, the procedure requires a definitional theory $Th := \truple{L}{Ax}{<}$, a set $ER$ of theory-specific rules for $Th$ and a set $M$ of minimal $ER$-proofs for $SF$. For every proof $\texttt{P}$ in $M$ and for every set $SF^\prime$ of signed formulas that is a proof-isomorphic candidate set for $SF$ with respect to $\texttt{P}$, we check if it is possible to construct an $ER$-proof for $SF^\prime$ that is deductively isomorphic to $\texttt{P}$ and add $SF^\prime$ in the output when this is the case.

The sets of signed formulas that result from running this procedure taking as input the set of signed formulas $SF_2:=\set{\Assert{x\in y\cap(w\cup z)}, \Deny{x\in (y\cap w)\cup z}}$, the set of expansion rules $TS_{sets}$ and the minimal $ER$-proofs for $SF_2$ are listed below:

\begin{itemize}
    \item $\set{\Assert{x\in y\cap(w\cup z)}, \Deny{x\in (y\cap w)\cup z}}$
    \item $\set{\Assert{x\in y\cap(w\symdiff z)}, \Deny{x\in (y\cap w)\cup z}}$
    \item $\set{\Assert{x\in y\setminus(w\setminus z)}, \Deny{x\in (y\setminus w)\cup z}}$
    \item $\set{\Assert{x\in y\setminus(w\symdiff z)}, \Deny{x\in (y\setminus w)\cup z}}$
    \item $\set{\Assert{x\in y\setminus(w\setminus z)}, \Deny{x\in (y\symdiff w)\cup z}}$
    \item $\set{\Assert{x\in y\setminus(w\symdiff z)}, \Deny{x\in (y\symdiff w)\cup z}}$
\end{itemize}

A minimal proof for the fourth set in the list is shown in Figure \ref{fig:search_proof_iso_min2}. Note that this proof is deductively isomorphic to the proof~$\texttt{M}$ of Figure \ref{fig:search_proof_iso_min}. In practical terms, this means that the proof exercises ``Prove that $x\in y\cap(w\cup z)$ implies $x\in (y\cap w)\cup z$\,'' and ``Prove that $x\in y\setminus(w\symdiff z)$ implies $x\in (y\setminus w)\cup z$\,'' exhibit comparable levels of proving complexity.

\begin{figure}[h]
\centering

\begin{forest}
[$\hypnode{1}{\Assert{x\in y\setminus(w\symdiff z)}}$ \\
    $\hypnode{2}{\Deny{x\in (y\setminus w)\cup z}}$ \\ 
        $\quad\node{3}{\Assert{x\in y}}{(1)}$ \\ 
            $\quad\node{4}{\Deny{x\in w\symdiff z}}{(1)}$ \\ 
                $\quad\node{5}{\Deny{x\in y\setminus w}}{(2)}$ \\ 
                    $\quad\node{6}{\Deny{x\in z}}{(2)}$ \\
                        $\quad\quad\;\,\node{7}{\Assert{x\in w}}{(3), (5)}$\\
                            $\quad\quad\;\,\node{8}{\Deny{x\in w}}{(4), (6)}$\\ $\quad\quad\absurd$ \\ $\quad\quad\close{(7) (8)}$, align=center
]
\end{forest}
\vspace{-1em}
\caption{A minimal proof for $\set{\Assert{x\in y\setminus(w\symdiff z)}, \Deny{x\in (y\setminus w)\cup z}}$.}
\label{fig:search_proof_iso_min2}
\end{figure}

\section{Conclusion}\label{sec:conclusion}

The present investigation has proposed a method for generating proof exercises with comparable levels of proving complexity. To the best of our knowledge, AQG tools incorporating mechanisms for difficulty control remain underexplored in the literature, particularly within the context of formal subjects. In our approach, difficulty control is grounded in the notion of proving complexity. To calculate the proving complexity of proof exercises, we developed theory-specific tableaux. The rules governing these tableaux are extracted from definitional axioms through a sequence of normal-form transformations. In addition, to formalize the notion of cut within theory-specific tableaux, we introduced an alternative notion of analyticity tailored to deductive systems free of logical symbols. The applicability of the proposed method was illustrated through an example based on a fragment of set theory.

As noted in Section~\ref{sec:Tsproofs}, certain definitional axioms do not permit the extraction of theory-specific rules due to the imposed analytic restrictions. In future work, we plan to investigate how these restrictions may be relaxed in order to accommodate a broader class of definitional axioms, while preserving termination of the proof-exercise generation procedure.
Another limitation, mentioned in Section~\ref{sec:complexity_proof_exercises}, is that proof exercises that cannot be logically described by a set into STSNF are out of our scope. Given a proof exercise $E$ such that the set $\Gamma$ is its logical description and $\Gamma$ cannot be converted into a set in STSNF, a solution could be to convert $\Gamma$ into a collection $C$ of sets of signed formulas $\Gamma_1$, $\dots$, $\Gamma_n$ in STSNF and use $C$ as the logical description of $E$. This solution would demand some adaptations in the method that we intend to investigate.

From a pedagogical perspective, the proving complexity of an exercise should remain invariant under permutations of its parameters or under the exchange of terms serving as arguments of certain specific symbols. For instance, let $\varphi^*$ be obtained from $\varphi$ by replacing $x \cap y$ with $y \cap x$. If $\{\Deny{\varphi}\}$ is refutable, then $\{\Deny{\varphi^*}\}$ is also refutable, and any minimal proof of $\{\Deny{\varphi}\}$ is deductively isomorphic to a minimal proof of $\{\Deny{\varphi^*}\}$. Consequently, the exercises $\{\Deny{p_1 \cap p_2 \subseteq p_1 \setminus p_2}\}$, $\{\Deny{p_2 \cap p_1 \subseteq p_1 \setminus p_2}\}$, and $\{\Deny{p_2 \cap p_1 \subseteq p_2 \setminus p_1}\}$ should be regarded as having comparable levels of proving complexity. In a continuation of the present work, we intend to formalize these invariance properties and incorporate them into the method, thereby enhancing its generative potential.

We also plan to conduct experiments in real pedagogical settings in order to validate our notion of proving complexity. In particular, we aim to investigate whether certain structural features of proofs are perceived as more challenging by students. For example, let $E_1$ and $E_2$ be two exercises with comparable levels of proving complexity. Suppose that $E_1$ admits a minimal proof composed exclusively of one-premise rules, whereas every minimal proof of $E_2$ involves at least one rule with multiple premises. It is plausible that students may experience greater difficulty solving $E_2$ than $E_1$, despite their comparable proving complexity as measured by our framework.

\printbibliography

\end{document}